\documentclass[10pt,a4paper]{article}
\usepackage{f1000_styles}
\usepackage{authblk}

\usepackage[numbers]{natbib}

\usepackage{graphicx}
\usepackage{txfonts}
\usepackage{natbib}
\bibpunct{(}{)}{;}{a}{}{,}
\input{bib.def}
\usepackage{xcolor,colortbl}

\definecolor{MyRed}{rgb}{0.9,0.0,0.0} 
\definecolor{MyDarkRed}{rgb}{0.6,0.0,0.0} 
\definecolor{MyLightRed}{rgb}{1.0,0.8,0.8} 
\definecolor{MyLightPink}{rgb}{1.0,0.9,0.9} 
\definecolor{MyLightOrange}{rgb}{1.0,0.8,0.5} 
\definecolor{MyPink}{rgb}{1.0,0.08,0.45} 
\definecolor{MyDarkBlue}{rgb}{0,0.08,0.45} 
\definecolor{MyDarkGreen}{rgb}{0,0.5,0.0} 
\definecolor{MyLightBlue}{rgb}{0.97,0.97,1.0} 
\definecolor{MyLightGreen}{rgb}{0.8,1.0,0.8} 
\definecolor{MyWarmYellow}{rgb}{1.0,0.95,0.85} 
\definecolor{MyMediumBlue}{rgb}{0.7,0.72,1.0} 

\usepackage[]{hyperref}
\hypersetup{
    colorlinks=true,
    linkcolor=blue,
    filecolor=magenta,      
    urlcolor=blue,
    citecolor=blue
}

\begin{document}
\pagestyle{fancy}

\title{Atacama Large Aperture Submillimeter Telescope (AtLAST) Science: Solar and stellar observations}


\author[1,2]{Sven Wedemeyer} 
\author[3]{Miroslav B\'{a}rta} 
\author[4]{Roman Braj\v sa}
\author[3]{Yi Chai}
\author[5]{Joaquim Costa}
\author[6]{Dale Gary} 
\author[5]{Guillermo Giménez de Castro}  
\author[3]{Stanislav Gun\'ar}
\author[6,7]{Gregory Fleishman} 
\author[8,9]{Antonio Hales }         
\author[10,11]{Hugh Hudson}   
\author[1,2]{Mats Kirkaune} 
\author[12,13]{Atul Mohan} 
\author[3,14]{Galina Motorina}
\author[15]{Alberto Pellizzoni}
\author[1,2]{Maryam Saberi} 
\author[6,16]{Caius L. Selhorst}
\author[5,10]{Paulo J. A. Sim\~oes}
\author[17,18]{Masumi Shimojo} 
\author[4]{Ivica Skoki\'c}
\author[4]{Davor Sudar}
\author[5]{Fabian Menezes}
\author[19]{Stephen White}
\author[20]{Mark Booth}
\author[20]{Pamela Klaassen}
\author[2]{Claudia Cicone}
\author[21]{Tony Mroczkowski}
\author[22]{Martin A. Cordiner}
\author[23-26]{Luca Di Mascolo}
\author[27,28]{Doug Johnstone}
\author[21]{Eelco van Kampen}
\author[29,30]{Minju Lee}
\author[31,32]{Daizhong Liu}
\author[33]{Thomas Maccarone}
\author[34]{John Orlowski-Scherer}
\author[35,36]{Am\'elie Saintonge}
\author[37]{Matthew Smith}
\author[38]{Alexander E. Thelen}

\affil[1]{Rosseland Centre for Solar Physics, University of Oslo, Postboks 1029 Blindern, N-0315 Oslo, Norway}
\affil[2]{Institute of Theoretical Astrophysics, University of Oslo, Postboks 1029 Blindern, N-0315 Oslo, Norway}
\affil[3]{Astronomical Institute, The Czech Academy of Sciences, 251 65 Ond\v rejov, Czech Republic}
\affil[4]{Hvar Observatory, Faculty of Geodesy, University of Zagreb, Ka\v{c}i\'ceva 26, HR-10000 Zagreb, Croatia}
\affil[5]{Centro de R\'adio Astronomia e Astrof\'isica Mackenzie, Escola de Engenharia, Universidade Presbiteriana Mackenzie, 01302-907, S\~ao Paulo, Brazil.}
\affil[6]{Center for Solar-Terrestrial Research, New Jersey Institute of Technology, Newark, NJ 07102, USA}
\affil[7]{Institute for Solar Physics (KIS),
Schoeneckstr 6, 79104 Freiburg,
Germany}
\affil[8]{National Radio Astronomy Observatory (NRAO), 520 Edgemont Road, Charlottesville, VA, 22903, USA}
\affil[9]{Joint ALMA Observatory (JAO), Alonso de Córdova 3107, Vitacura, 763-0355, Santiago, Chile} 
\affil[10]{SUPA School of Physics and Astronomy, University of Glasgow, Glasgow G12 8QQ, UK}
\affil[11]{Space Sciences Laboratory, University of California, Berkeley, Berkeley, CA, 94720, USA}
\affil[12]{The Catholic University of America, Washington, DC 20064, USA}
\affil[13]{NASA Goddard Space Flight Center, Greenbelt, MD 20771, USA}
\affil[14]{Central  Astronomical Observatory at Pulkovo of Russian Academy of Sciences, St. Petersburg, 196140, Russia}
\affil[15]{INAF - Cagliari Astronomical Observatory, Via della Scienza 5, I--09047 Selargius (CA), Italy}
\affil[16]{NAT - N\'ucleo de Astrof\'isica, Universidade Cidade de S\~ao Paulo, S\~ao Paulo, SP, Brazil}
\affil[17]{National Astronomical Observatory of Japan, Mitaka, Tokyo, 181-8588, Japan}
\affil[18]{Graduate University for Advanced Studies, SOKENDAI, Mitaka, Tokyo, 181-8588, Japan}
\affil[19]{Space Vehicles Directorate, Air Force Research Laboratory, Albuquerque, NM 87117 USA}
\affil[20]{UK Astronomy Technology Centre, Royal Observatory Edinburgh, Blackford Hill, Edinburgh EH9 3HJ, UK}
\affil[21]{European Southern Observatory, Karl-Schwarzschild-Str. 2, 85748 Garching bei M\"unchen, Germany}
\affil[22]{Astrochemistry Laboratory, Code 691, NASA Goddard Space Flight Center, Greenbelt, MD 20771, USA.}
\affil[23]{Laboratoire Lagrange, Université Côte d'Azur, Observatoire de la Côte d'Azur, CNRS, Blvd de l'Observatoire, CS 34229, 06304 Nice cedex 4, France}
\affil[24]{Astronomy Unit, Department of Physics, University of Trieste, via Tiepolo 11, Trieste 34131, Italy}
\affil[25]{INAF -- Osservatorio Astronomico di Trieste, via Tiepolo 11, Trieste 34131, Italy}
\affil[26]{IFPU -- Institute for Fundamental Physics of the Universe, Via Beirut 2, 34014 Trieste, Italy}
\affil[27]{NRC Herzberg Astronomy and Astrophysics, 5071 West Saanich Rd, Victoria, BC, V9E 2E7, Canada}
\affil[28]{Department of Physics and Astronomy, University of Victoria, Victoria, BC, V8P 5C2, Canada}
\affil[29]{Cosmic Dawn Center (DAWN), Denmark}
\affil[30]{DTU-Space, Technical University of Denmark, Elektrovej 327, DK2800 Kgs. Lyngby, Denmark}
\affil[31]{Max-Planck-Institut f\"{u}r extraterrestrische Physik, Giessenbachstrasse 1 Garching, Bayern, D-85748, Germany}
\affil[32]{Purple Mountain Observatory, Chinese Academy of Sciences, 10 Yuanhua Road, Nanjing 210023, China}
\affil[33]{Department of Physics \& Astronomy, Texas Tech University, Box 41051, Lubbock TX, 79409-1051, USA}
\affil[34]{Department of Physics and Astronomy, University of Pennsylvania, 209 South 33rd Street, Philadelphia, PA, 19104, USA}
\affil[35]{Department of Physics and Astronomy, University College London, Gower Street, London WC1E 6BT, UK}
\affil[36]{Max-Planck-Institut f\"ur Radioastronomie (MPIfR), Auf dem H\"ugel 69, D-53121 Bonn, Germany}
\affil[37]{School of Physics \& Astronomy, Cardiff University, The Parade, Cardiff CF24 3AA, UK}
\affil[38]{Division of Geological and Planetary Sciences, California Institute of Technology, Pasadena, CA 91125, USA.}

\maketitle
\thispagestyle{fancy}


\begin{abstract}
Observations at (sub-)millimeter wavelengths offer a complementary perspective on our Sun and other stars, offering significant insights into both the thermal and magnetic composition of their chromospheres. Despite the fundamental progress in (sub-)millimeter observations of the Sun, some important aspects require diagnostic capabilities that are not offered by existing observatories. In particular, simultaneous observations of the radiation continuum across an extended frequency range would facilitate the mapping of different layers and thus ultimately the 3D structure of the solar atmosphere. Mapping large regions on the Sun or even the whole solar disk at a very high temporal cadence would be crucial for systematically detecting and following the temporal evolution of flares, while synoptic observations, i.e., daily maps, over periods of years would provide an unprecedented view of the solar activity cycle in this wavelength regime. As our Sun is a fundamental reference for studying the atmospheres of active main sequence stars, observing the Sun and other stars with the same instrument would unlock the enormous diagnostic potential for understanding stellar activity and its impact on exoplanets. The Atacama Large Aperture Submillimeter Telescope (AtLAST), a single-dish telescope with 50m aperture proposed to be built in the Atacama desert in Chile, would be able to provide these observational capabilities. Equipped with a large number of detector elements for probing the radiation continuum across a wide frequency range, AtLAST would address a wide range of scientific topics including the thermal structure and heating of the solar chromosphere, flares and prominences, and the solar activity cycle. In this white paper, the key science cases and their technical requirements for AtLAST are discussed.
\end{abstract}

\section*{\color{OREblue}Keywords}


%
Sun: activity; 
Sun: atmosphere; 
Sun: filaments, prominences; 
Sun: flares; 
Sun: magnetic fields; 
Sun: solar-terrestrial relations; 
Sun: sunspots;

\section*{Plain language summary}
Observations of our Sun and other stars at wavelengths of around one millimeter, i.e. in the range between infrared and radio waves, present a valuable complementary perspective. Despite significant technological advancements, certain critical aspects necessitate diagnostic capabilities not offered by current observatories. The proposed Atacama Large Aperture Submillimeter Telescope (AtLAST), featuring a 50-meter aperture and slated for construction at a high altitude in Chile’s Atacama desert, promises to address these observational needs. Equipped with novel detectors that would cover a wide frequency range, AtLAST could unlock a plethora of scientific studies contributing to a better understanding of our host star. 
Simultaneous observations over a broad frequency range at rapid succession would enable the imaging of different layers of the Sun, thus elucidating the three-dimensional thermal and magnetic structure of the solar atmosphere and providing important clues for many long-standing central questions such as how the outermost layers of the Sun are heated to very high temperatures, the nature of large-scale structures like prominences, and how flares and coronal mass ejections, i.e. enormous eruptions, are produced. The latter is of particular interest to modern society due to the potentially devastating impact on the technological infrastructure we depend on today. Another unique possibility would be to study the Sun’s long-term evolution in this wavelength range, which would yield important insights into its activity cycle. Moreover, the Sun serves as a fundamental reference for other stars as, due to its proximity, it is the only star that can be investigated in such detail. The results for the Sun would therefore have direct implications for understanding other stars and their impact on exoplanets.  
This article outlines the key scientific objectives and technical requirements for solar observations with AtLAST.

\pagestyle{fancy}
\twocolumn

\section{Introduction}
The continuum radiation emitted by the Sun at millimeter wavelengths originates from the chromosphere, the layer of the solar atmosphere located between the photosphere (i.e., the Sun’s surface layer) and the corona. The thermal and magnetic structure of the chromosphere is complex and highly dynamic with a plethora of physical processes at work \citep[e.g.,][]{2007Sci...318.1574D,2007PASJ...59S.655D,Jess_2009Sci...323.1582J,2012Natur.486..505W,2014Sci...346B.315T,2020SSRv..216..140V}
 many of which induce notable deviations from equilibrium conditions \citep[for example in the ionisation state of hydrogen,][]{2002ApJ...572..626C,2024RSPTA.38230230H}. Next to the chromospheric plasma, the continuum at millimeter wavelengths can also probe the cool, dense coronal plasma in filaments and prominences, thus accessing a large range of spatial scales. The largest features, like filaments can span substantial fractions of the solar diameter (($\sim$30'), followed by Active Regions (often a few arcminutes), and magnetic network cells as the chromospheric imprints of supergranulation cells (20” – 60”). Sunspots \citep{2003A&ARv..11..153S},  
including the penumbra, have diameters of typically 30'' but occasionally up to $\sim1$', whereas the atmosphere in the Quiet Sun regions exhibits spatial scales of 1-2'' and below. It is expected that there is small-scale structure even smaller than $\sim0.1$'' scales resolved by modern 4m-class optical solar telescopes 
\citep{2020SoPh..295..172R, 2022A&A...666A..21Q}. Likewise, the temporal scales range from about a month for one solar rotation, days for the evolution of sunspots, down to minutes and seconds for small-scale and energetic phenomena. 

Consequently, observing the chromosphere and characterising the physical mechanisms that govern this dynamic environment is challenging. 
So far, there were only a few suitable diagnostic tools available to examine the plasma conditions in the chromosphere such as the spectral lines of singly ionized calcium (Ca\,II) and magnesium (Mg\,II) as well as the H$\alpha$  line. The interpretation of observations in these spectral lines and deriving the physical properties of the plasma in the mapped atmospheric regions is complicated and limited by the complex line formation mechanisms that involve non-local thermodynamic equilibrium effects. The situation is better for the millimeter continuum radiation, which is assumed to be formed mostly under local equilibrium conditions so that the measured brightness temperature should be closely related to the local plasma temperature. Although available telescopes for millimeter wavelengths made notable contributions to solar physics in the past \citep[e.g.,][and references therein]{1959AnAp...22....1K,2011SoPh..273..339T,2014A&A...561A.133L}, they did not have sufficient spatial resolution for resolving the small spatial scales in the solar chromosphere until the Atacama Large Millimeter/submillimeter Array (ALMA) started regular observations of the Sun in 2016 \citep{2002AN....323..271B,2011SoPh..268..165K,2017SoPh..292...87S,2017SoPh..292...88W}. It should also be noted that the chromosphere evolves on timescales well below 1\,min, which means that meaningful observations are restricted to very short integration times, essentially leading to snapshot imaging. 
Techniques like Earth rotation synthesis or combining observations in different array configurations are not viable for solar observations.

\begin{figure*}[t]
    \centering
    \includegraphics{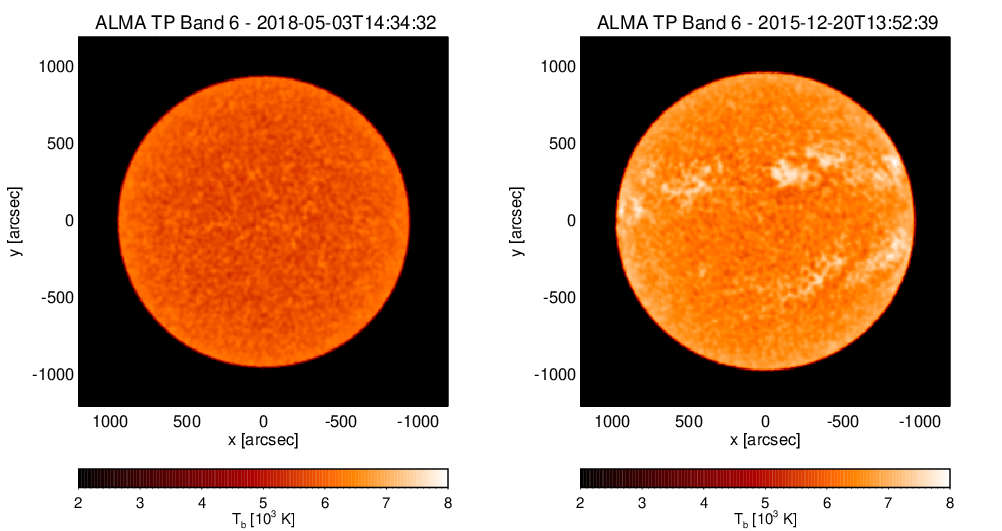}
    \caption{ALMA Total Power (TP) maps in Band~6 (230\,GHz) obtained by double-circle scanning at two different times capturing a very quiescent state (left) and a more active state with clearly visible Active Regions including sunspots (right). 
    (ALMA project IDs: 2017.1.00009.S, 2011.0.00020.SV.)}  
    \label{fig:almasdo}
\end{figure*}

The high spatial and temporal resolution and also the prospects for utilising polarisation measurements for measuring magnetic fields in the solar chromosphere make (sub-)millimeter telescopes 
very powerful tools for investigating the solar chromosphere and thus contributing to the solution of many open questions 
\citep[see][and references therein]{2016SSRv..200....1W}. 

It should also be noted that the continuum radiation received at a certain frequency mostly originates from a narrow layer in the solar atmosphere with the mapped height range depending on the used receiver band. 
Observations of the Sun in different receiver bands thus map different layers in the Sun's atmosphere. 
Despite already large success of ALMA in advancing solar physics regarding a broad range of topics from the general thermal stratification of the solar atmosphere, large-scale structures like prominences to the energetics of small-scale events like shock waves and spicules \citep[e.g.][and many more]{2017A&A...605A..78A,Shimojo2017ApJ...841L...5S,2018A&A...620A.124D,2020A&A...634A..56D,2020A&A...644A.152E,roman2018A&A...613A..17B,Rodger2019ApJ...875..163R,2019ApJ...881...99M,Selhorst2019ApJ...871...45S,2021ApJ...906...82C,ValleSilva2021MNRAS.500.1964V,Labrosse2022MNRAS.513L..30L,2022ApJ...927L..29H,Menezes2022MNRAS.511..877M,Oliveira2022,Skokic2023A&A...669A.156S}, there are important science cases that require different observational capabilities. 
 Despite already large success of ALMA in advancing solar physics,  there are important science cases that require different observational capabilities. 
One diagnostic limitation is that observations are limited to one narrow frequency band and therefore to one atmospheric layer at a time, while the solar atmosphere is a prime example of a dynamic, complex, and inherently 3D phenomenon. 
Therefore, meaningful observations must be carried out strictly simultaneously with as many complementary diagnostics and in as many layers as possible, which is also required to facilitate analysis techniques such as spectroscopic inversions and differential emission measures \citep[e.g.,][]{2016SSRv..200....1W,2018A&A...620A.124D,2020A&A...634A..56D,2020A&A...643A..41D,2022FrASS...9.8249A,2022ApJ...933..244H,2022FrASS...925368L}. 
This capability could be offered by a large-aperture sub-millimeter facility telescope like the proposed Atacama Large Submillimeter
 Telescope \citep[AtLAST\footnote{\url{http://atlast-telescope.org/}},][]{2020SPIE11445E..2FK,2022SPIE12190E..07R,2023ursi.confE.174M,Mroczkowski2024}. 
Despite an impressive aperture of 50\,m, AtLAST would not reach the angular resolution achieved with an interferometric array like ALMA but compensate with other unique features such as a simultaneous wider frequency coverage, thus unlocking complementary science cases.  
AtLAST would thus have the potential to boost the impact of millimeter observations on solar physics beyond its current state. 
In this article, key science cases and observing strategies for solar observations with AtLAST are described. 
At AtLAST’s current stage in the development process, the general capabilities of the telescope such as diameter field-of-view, mapping speed etc. have been decided \citep{Mroczkowski2024}, while the technical aspects of the up to 6 foreseen instruments remain open. Please note that the purpose of this article is to set out the technical requirements placed on the telescope, its instruments, and its operations that are necessary to achieve great advances in solar physics  \citep[please see][for a summary of all science cases]{2024SPIE13102E..06B}. The discussion of the solar science cases presented here aims at defining the corresponding technical requirements. 
 
\section{Key science cases} 
\label{sec:keyscience}

The chromosphere is an integral part of the solar atmosphere and as such plays an important role in the transport of energy and matter. These are essential for understanding the heating of the corona, the origins of the solar wind, and the drivers of solar activity and space weather. There is a plethora of physical processes involved, which are entangled in complicated dynamic ways, rendering the understanding of the workings of our Sun a challenging undertaking. 

As detailed in the sections below, AtLAST \citep{Mroczkowski2024} opens up a new window to science cases that cannot be addressed with existing sub-mm telescopes. 
In addition, some solar science cases for ALMA \citep[see][and references therein; see also Fig.~\ref{fig:almasdo} for examples of ALMA single-dish scans of the Sun]{2016SSRv..200....1W,2018Msngr.171...25B} and LLAMA \citep{llama:2021} like, e.g., atmospheric heating, flares, and prominences, are potential solar use cases for AtLAST, too, thus creating synergies between these observatories. These science cases concern the thermal structure of the solar atmosphere including the thermal structure of prominences and filaments, the related transport of energy and mass, and phenomena related to solar activity. 
In the following subsections, the key science cases for solar observations with AtLAST are highlighted. 
Please note that, at this stage, the telescope concept design and requirements are well-defined \citep{Mroczkowski2024}.  This includes the technical requirements such as the optical design, receiver cabin layout, primary aperture diameter (50\,m), scan speed (up to 3\,deg/s), field of view, approximate frequency range ($\sim30 - 950$\,GHz, corresponding to 0.3 -- 10\,mm), and surface accuracy goals, along with several other key technological concepts such as approaches for metrology, instrument exchange, and energy recovery. However, the first-generation instrumentation priorities have not yet been determined and are under consideration using the solar case presented and other science cases.  The resulting instrumental requirements are discussed in Section~\ref{sec:technical}. 

\subsection{Thermal structure and the atmospheric heating problem}
\label{sec:coronalheating}

The chromosphere plays an important role in the transport of energy and matter from the vast energy reservoir in the Sun’s interior underneath the photosphere to the transition region and corona above. This transport phenomenon and the chromosphere’s role in it are essential for understanding the heating of the corona, the origins of the solar wind, and the drivers of solar activity and space weather. The boundaries between the chromosphere and the layers below and above, however, vary strongly in time and space, making it impossible to define strict height ranges for the layers. Rather, the chromosphere must be understood as an integral part of the solar atmosphere \citep{2009SSRv..144..317W}. The chromosphere can be more suitably characterised based on the notable changes in physical behaviour including deviations from equilibrium conditions and the transition from a thermodynamically to a magnetically dominated domain as expressed by the corresponding change of the ratio of thermal to magnetic pressure, the so-called plasma beta. These changes also give rise to a plethora of physical processes, which are entangled in complicated dynamic ways, rendering the understanding of the workings of our Sun a challenging undertaking \citep{2019ARA&A..57..189C}.

One of the large open questions in solar physics is the coronal heating problem, which has been known since the late 1930s when observations of spectral lines due to elements in extremely high ionisation stages implied temperatures in excess of a million Kelvin in the corona \citep{1943ZA.....22...30E}. 
As pointed out in the introduction, deriving the plasma properties in the outer solar atmosphere is difficult, which hampered identifying the processes responsible for the transport of the required energy into the corona. As a result, despite significant progress 
\citep[e.g.,][ and many more]{2004ApJ...617L..85P,2005ApJ...618.1031G,2006SoPh..234...41K,2011Natur.475..477M}, much of the coronal heating problem remains unsolved today. It should be noted that the Sun is not unique in this sense but rather serves as a reference for other stars that exhibit a corona, too, thus making the solar coronal heating problem relevant for all stars with notable surface convection, i.e. in particular main sequence stars of spectral types G-M.  

After decades of research, numerous processes have been identified that are capable of supplying the energy required to account for the high temperatures observed in the Sun. The focus has shifted towards determining which of these processes are most relevant and how exactly they transport and deposit energy. It is likely that a combination of different processes contributes to chromospheric/coronal heating, with their relative importance varying in different solar regions characterized by distinct magnetic field environments and activity levels, i.e. from Active Regions to Quiet Sun regions. Some processes provide continuous heating, while others, such as solar flares, produce a more variable and intermittent heating component due to their transient nature.

Typically these phenomena are grouped into processes related to (i)~magnetic reconnection/Ohmic heating and (ii)~wave heating processes, specifically acoustic and magnetohydrodynamic (MHD) waves. MHD waves, including Alfvén waves \citep{2007Sci...318.1574D,Jess_2009Sci...323.1582J,2010ApJ...711..164V,2014Sci...346D.315D}, can contribute to plasma heating by perturbing the magnetic field, leading to wave damping and the release of magnetic and kinetic energy,  but, in turn, can also be excited by dynamic variations of magnetic field structures. 
There is a plethora of physical processes and effects that are not only relevant for observations of the Sun but also in a wider plasma physics context, e.g., multi-fluid effects \citep{2010ApJ...724.1542K,2012ApJ...753..161M}, plasma instabilities \citep[e.g.,][]{2008A&A...480..839F}, and turbulence \citep{2010SSRv..156..135P}. See the review on ALMA’s potential for addressing the chromospheric/\linebreak{}coronal heating problem by \citet{2016SSRv..200....1W}, which is equally relevant for AtLAST. 
The resulting heating is evident in phenomena like spicules observed at the solar limb.  The multitude of possible heating mechanisms and their potential entanglement has made it challenging to determine which ones are predominant and how their contributions depend on the solar region type. In principle, identifying these mechanisms and understanding their contributions requires quantitative and precise measurements that capture the thermal, magnetic, and kinetic state of chromospheric plasma over time and in three spatial dimensions. AtLAST has the potential to provide such data.  

Another important aspect concerns the formation heights of the solar (sub)mm continuum, i.e., the exact layers from where the continuum radiation at a given frequency emerges. ALMA observations and corresponding simulations have led to surprising findings like the imprint of coronal loops in Band~3 (3\,mm) observations   \citep[see, e.g., ][]{2020A&A...635A..71W} and a strong variance of the relative formation heights of Band 3 and Band 6  \citep[see, e.g.,][]{2021ApJ...906...83C,2021RSPTA.37900185E}. 
See \citep{2022FrASS...967878W} for a discussion. Strictly simultaneous observations across a wide frequency range --  and consequently a broad range of formation heights --  could enable AtLAST to better constrain the formation heights and thus allow a more reliable interpretation and reconstruction of the atmospheric 3D structure.

\subsection{Solar flares}
\label{sec:flares}

AtLAST is poised to make significant contributions in addressing numerous unresolved inquiries pertaining to solar flares, a prominent conundrum within contemporary solar physics that continues to be a highly dynamic area of investigation \citep[see, e.g.,][and references therein for more details]{2016SSRv..200....1W}.
Although numerous intricacies remain shrouded in mystery, it is evident that solar flares manifest as a result of the dynamic reconfiguration and interconnection of magnetic fields within the solar atmosphere 
\citep[see, e.g.,][]{1994ApJ...424..436W,2000ApJ...531L..75H,2012ApJ...748...77S,2015A&A...581A...8R,2019ApJS..240...11P}. 
Consequently, substantial amounts of energy previously stored within the magnetic field are explosively released during these events, manifesting as both radiation and high-energy particles. This emitted radiation encompasses the entire electromagnetic spectrum, spanning from gamma and X-rays to mm and radio waves. 

Solar flares can differ by orders of magnitude in strength, which is typically measured according to the emitted soft X-ray flux. The strongest solar flares ever documented released energy on the order of $10^{32}$\,ergs 
 \citep{2005JGRA..11011103E}  
and are often accompanied by coronal mass ejections (CME) that can impact space weather with notable effects on Earth. The effects manifest themselves from beautiful aurorae to severe disturbances of power grids (like the Quebec blackout in 1989) and satellite infrastructure. As a recent example, we can mention the destruction of the majority of a batch of Starlink satellites due to solar activity \citep[e.g.][]{Dang2022SpWea..2003152D}. And yet, it cannot be ruled out that even stronger flares exceeding the Carrington event of 1859 could occur with disastrous consequences for our technology-dependent modern society. 
Such so-called super-flares are observed for other (solar-like) stars \citep{2012Natur.485..478M} (see Sect.~\ref{sec:stellar}). 
While the bolometric radiative energy release of the Carrington event is estimated to be on the order of $\sim5 \times 10^{32}$\,erg 
\citep[or $2 \times 10^{33}$\,erg when adding the kinetic energy of a CME; ][]{2013JSWSC...3A..31C}
 while TESS observations exhibit stellar flares with radiative energies in excess of 10$^{35}$\,erg \citep{2022ApJ...935..143P}. The study of strong solar flares is thus of utmost importance. An essential observational challenge, which is hampered fully revealing the physical mechanisms behind flares, is connected to the small field of view of many solar telescopes, making it hard to point at the right position at the right time to capture the early stages of a flare. High-cadence observations of the whole solar disk (or at least large regions) with AtLAST (see Sect.~\ref{sec:spatialmap}) would thus increase the number of fully observed flares, contributing to unveiling the physical mechanisms behind flares in new ways.

The occurrence rate of a flare is roughly related to the released energy, making the strongest events rare (although varying with the solar cycle) and the much weaker micro- and nano-flares more frequent and ubiquitous. The latter are therefore extremely important as potential contributions to the heating of the corona (see Sect.~\ref{sec:coronalheating}) in a more continuous way across most of the Sun, whereas strong flares are associated with Active Regions. 

Despite many decades of research and significant progress, there are still many open questions regarding central aspects of solar and stellar flares on all scales including the particle acceleration mechanisms \citep[e.g.][]{Miller1998SSRv...86...79M}, quasi-periodic pulsations \citep[e.g.][]{Simoes2015SoPh..290.3625S,2021ApJ...909L...1M}, and the source of the still enigmatic emission at sub-THz frequencies 
\citep{Kaufmann1985STIN...8616178K,Kaufmann2004ApJ...603L.121K,Fl_Kontar_2010, Krucker2013A&ARv..21...58K}, i.e. in the frequency range to be covered by AtLAST. 

\begin{figure}[t]
    \centering
    \includegraphics[width=0.5\textwidth]{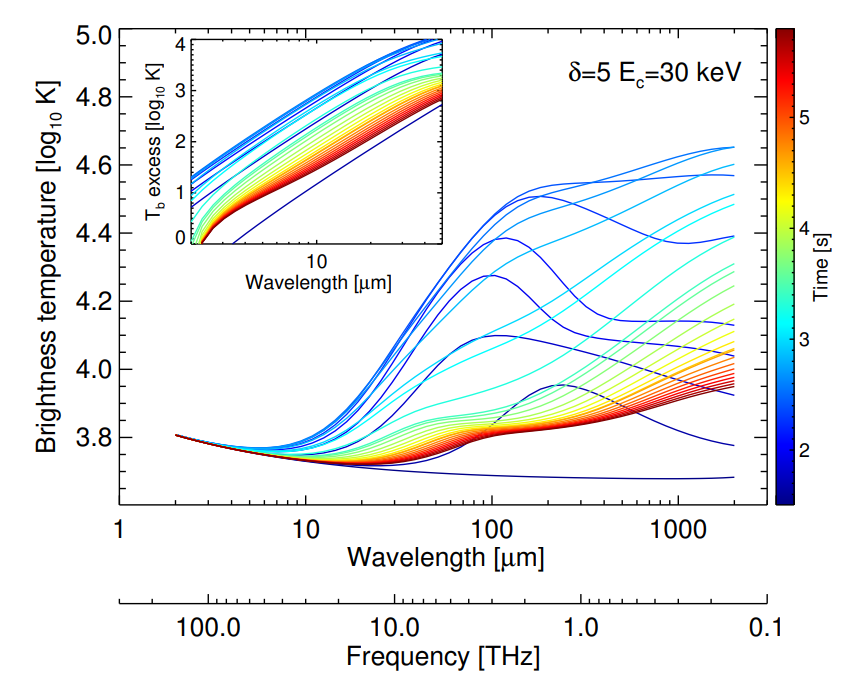}
    \caption{Evolution of a synthetic sub-mm and IR flare $T_b$ spectrum from RHD modelling \citep[from ][]{Simoes2017A&A...605A.125S}. 	From these simulations, for $\lambda > 50 \mu$m, $T_b$ should provide the electron temperature in the optically thick upper chromosphere \citep[see, e.g.,][and references therein]{2016SSRv..200....1W}. Please note that the range relevant for AtLAST is for frequencies less than $\sim$1.0\,THz. 
\label{fig:synthspectra}}
\end{figure}

Observations of flare-like brightenings with ALMA, in both interferometric mode  \citep{Shimojo2017ApJ...841L...5S,Rodger2019ApJ...875..163R} and single-dish \citep{Skokic2023A&A...669A.156S}, have revealed faint and localized sources, likely thermal in origin, and well correlated  with the observed emissions in EUV and SXR. Observations of such sub-mm flares might also provide key information about the formation of hot onsets in solar flares 
revealing the increase of the electron temperature of the chromosphere (see Figure~\ref{fig:synthspectra}) or the presence of non-thermal electrons during flare onsets \citep{hudson2021MNRAS.501.1273H,douglas2023MNRAS.525.4143D}. 

The mm and sub-mm data can uniquely address several key questions regarding solar flares  \citep{2022FrASS...9.6444F}.
For example, in combination with other instruments for higher frequencies (see Sect.~\ref{sec:synergy}), the spectral coverage in the mid-IR to sub-mm range can help to unveil the thermal structure of the flaring atmosphere. Radiative-hydrodynamic (RHD) modelling of flares suggest that the emission at wavelengths longer than 50~$\mu$m is optically thick, and thus, the observed brightness temperature should directly provide estimates for the electron temperature at the emitting chromospheric layer \citep{Simoes2017A&A...605A.125S}. In Figure~\ref{fig:synthspectra}, we provide an example of the temporal evolution of a synthetic flare spectrum from RHD modelling. 


Addressing these aspects with AtLAST would require high spectral, spatial and temporal resolution and preferably full polarisation (see Sect.~\ref{sec:technical}), which then would allow for probing the dynamic thermal structure and magnetic structure of the solar atmosphere before, during, and after a flare with potentially ground-breaking insights.  While already the thermal free-free radiation produced mostly due by electron-ion free-free absorption and H$^-$ free-free absorption \citep[e.g.][]{Heinzel2012SoPh..277...31H} contains essential information about the physical mechanisms behind flares, the non-thermal radiation component due to gyrosynchrotron and resonance processes would be essential for constraining the acceleration of charged particles, which is central to understanding flares. 

\subsection{Solar prominences} 

Multi-scale observations of solar prominences, i.e.  extended structures in the solar atmosphere that stretch upward from the visible surface into the corona (Vial \& Engvold 2015), are another area where AtLAST can make a significant impact. The magnetic field in a prominence supports denser prominence plasma against gravity. The prominence material has a much lower temperature (around 10$^4$\,K) while being surrounded by coronal plasma (around $10^6$\,K). Here, the prominence magn
etic field provides the insulation, which allows the prominence plasma to cool radiatively. Prominences appear as bright when viewed above the solar limb but dark when viewed against the solar disk's brightness (when they are called filaments).  Quiescent prominences can extend over a few hundred thousand kilometers, i.e. a significant fraction of the Sun's diameter. These prominences are among the longest-lasting solar phenomena with lifetimes of several days to weeks. 
In contrast, erupting prominences, as one of the main drivers of space weather, are capable of hurling large amounts of magnetized plasma into interplanetary space at high speeds. The large field of view of AtLAST (45'' at 950\,GHz to 450'' at 100\,GHz), combined with its good spatial resolution (1.5'' at 950\,GHz) and the ability to observe the Sun once a day for extended periods, will offer us opportunities to study the questions of the prominence origin and evolution, together with the drivers that make the prominence magnetic field configurations unstable and erupt. 

Moreover, AtLAST will provide simultaneous multi-band coverage, which is a capability that no other facility can provide, at least not in combination with the required angular resolution. AtLAST will therefore enable us to answer also important questions about the thermal properties and energy balance of prominence plasma and to uncover the nature of the prominence plasma transition region spanning from cool cores to the hot corona. Another advantage of observing at millimetre wavelengths lies in their easier interpretation. The spectral lines in the optical and UV used for prominence observations are typically optically thick so that detailed radiative transfer calculations are necessary for their analysis. In contrast, the prominence plasma is mostly optically thin at AtLAST wavelengths, which makes the interpretation much simpler due to the relation between the observed flux and the plasma temperature of the emitting region \citep[see, e.g.,][]{Rodger2017SoPh..292..130R,2016ApJ...833..141G,2018ApJ...853...21G}. Moreover, simultaneous observations in multiple bands, which only AtLAST can provide, allow detailed measurements of the kinetic temperature distribution of the prominence plasma \citep[see][]{2016ApJ...833..141G,2018ApJ...853...21G}.
Nevertheless, coordinated observations at millimetre wavelengths by AtLAST and in optical spectral lines (such as H$\alpha$) and UV lines (e.g., Mg II h$\&$k) would bring large benefits. This was demonstrated by, for example, \citet{2018A&A...620A.124D} and \citet{2022ApJ...933..244H}. Combined observations of prominences in the H$\alpha$ line and at millimetre wavelengths (by ALMA) were used to derive the temperature structure of prominence plasma by \citet{Labrosse2022MNRAS.513L..30L} and \citet{2022ApJ...927L..29H}. Similar coordinated observations of the solar chromosphere were used by \citet{2020A&A...634A..56D}  and \citet{2021ApJ...922..113S} and in Active Regions by \citet{2020A&A...643A..41D}  and \citet{2022FrASS...925368L}.

Joint observations of prominences with AtLAST and ALMA will also be extremely valuable. ALMA interferometry is capable of very high spatial resolution, thus extending the observable domain towards the shortest resolvable scales. On the other hand, the interferometric (Fourier-based) observations are insensitive to the larger-scale structures (larger than the single antenna primary beam). In order to fill the gap around zero in the Fourier space, ALMA uses supplementary single-dish (SD) scanning when observing a multi-scale structures, like prominences. However, the ALMA SDs have the same size of 12\,m and, consequently, there is a little overlap between the domains of scales covered by the high-resolution interferometric data and those obtained from the SD observations. With the much larger dish, AtLAST would provide an excellent bridge connecting both domains and allowing for a proper match of the interferometric and SD data. This is very true in general, but for the limb observations (prominences, spicules) even yet more vital. 

\subsection{The solar activity cycle}
\label{sec:solcycle}
As stated above, millimetre observations have the advantage that the brightness temperatures corresponding to the observed continuum flux are thought to be a close proxy for the chromospheric plasma temperature with the formation height increasing with wavelength. Measuring the polarisation allows for deriving the line-of-sight component of the magnetic field. 
Daily full-disk maps of the Sun would permit the study of how the temperature (and hence the energy content) evolves in Active Regions, Quiet Sun regions and coronal holes. These data can be compared with other diagnostics \citep{2017ApJ...835...25E}, and in particular the evolution of magnetic fields in the solar atmosphere \citep[see, e.g.,][]{2012ApJ...757L...8L,2017SoPh..292...73T} for the relation of magnetic field measurements and the radio flux at a wavelength of 10.7\,cm). The solar chromosphere is responsible for most of the UV emission that plays a major role in structuring the Earth’s upper atmosphere, and evolution of the millimeter emission will in turn reflect the drivers of solar UV emission. As observations in different frequency bands probe different depths in the solar atmosphere \citep{1981ApJS...45..635V}, comparing the evolution in these different layers is important for understanding the transport of energy through the chromosphere. While the solar cycle is already intensively studied at other wavelengths \citep[see][and references therein]{2015LRSP...12....4H}, repeated sets of such observations obtained over a period of years will reveal how the millimeter emission responds to the solar cycle. A noteworthy demonstration of the scientific potential is the sequence of observations with the Solar Submillimeter-wave Telescope \citep[SST, ][]{2008SPIE.7012E..0LK} that covers a period longer than one solar cycle \citep{Menezes2021ApJ...910...77M}. 

Measuring the solar radius at subterahertz frequencies \linebreak (submm/mm) allows one to probe the solar atmosphere, since these measurements show the height above the photosphere at which most of the emission at a determined observation frequency is generated \citep{Menezes2017SoPh..292..195M}.
Changes in the solar radius show it can be modulated with the 11-year activity cycle (mid-term variations) as well as longer periods (long-term variations). 
The equatorial radius time series was found to be positively correlated to the solar cycle, since the equatorial regions are more affected by the increase of active regions during solar maxima, making the solar atmosphere warmer in these regions; on the other hand, the anticorrelation between polar radius time series and the solar activity proxies could be explained by a possible increase of polar limb brightening during solar minima \citep{Menezes2021ApJ...910...77M}.
In the context of the study on the subterahertz solar atmosphere, spicules can affect the measurement of the solar radius and cause the solar limb to appear more diffuse, which can lead to an overestimation of the solar radius. 
In \citet{Menezes2022MNRAS.511..877M}, the discrepancies between measured limb brightening values and model predictions highlight the need for further studies to improve our understanding of the solar atmosphere.
Therefore, it is important to take into account the presence of spicules when measuring the solar radius and limb brightening at different frequencies. 
Currently, however, there are no full-disk observations of the Sun in the mm range apart from occasional ALMA single-dish Total Power (TP)  maps a few times a year.

\begin{figure*}[t]
    \centering
    \includegraphics{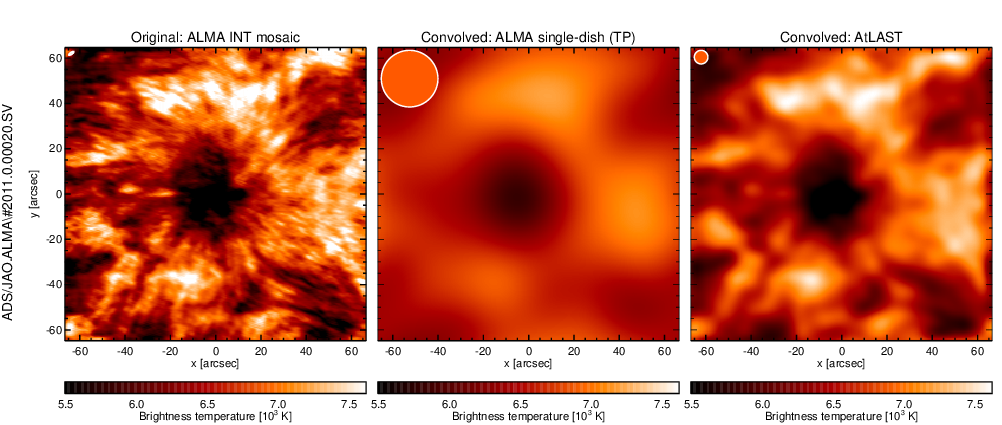} 
    \caption{The ALMA sunspot mosaic (left) that was taken during the ALMA SV campaign in on December 18, 2015 in Band~6 230\,GHz at original interferometric resolution (left) and after convolution with a beam corresponding to a single ALMA 12-m (Total Power) antenna (middle) and after convolution with the expected AtLAST beam (right).  For reference, the corresponding beam is plotted in the upper left corner of each panel. } 
    \label{fig:sunspot}
\end{figure*}

Please note that the Sun’s rotation depends on latitude with a (sidereal) rotation period of down to ~24 days at the equator and up to 35 days at the poles \citep[see, e.g.,][]{1983ApJ...270..288S,1986A&A...155...87B,1998ApJ...505..390S}. Any long-lived feature like an Active Region would thus move across the visible disk of the Sun in less than two weeks. However, at the resolution anticipated for AtLAST (see Figure~\ref{fig:sunspot}), features like Active Regions and sunspots \citep{2003A&ARv..11..153S}  therein would evolve significantly on a daily timescale. The evolution of Active Regions and the change of the Sun’s magnetic nature across a solar cycle would therefore be addressed best via full-disk mapping of the Sun at multiple frequency bands, once per day. Already an initial 1-month campaign covering a full solar rotation would allow for addressing the fundamental unsolved problems mentioned above including the nature of coronal and chromospheric heating problems, whereas much longer sequences are required for addressing the solar activity cycle. Such sequences would also be relevant for producing a larger sample of flare observations, where the individual observations should be time series over an hour or more as daily maps would not capture the temporal evolution of individual flares (see Sect.~\ref{sec:flares}). In addition, these sequences would provide vital complementary input for predictions of the solar activity cycle, which yield important societal impact but, despite much progress, remain challenging \citep[e.g., ][]{1999JGR...10422375H,2006GeoRL..33.5102D,2007ApJ...659..801C,2007PhRvL..98m1103C,2012SoPh..281..507P,2014SSRv..186....1B}. 

New insights regarding the Sun’s activity variations and thus the long-term evolution of our host star could be transferred to other stars as demonstrated by stellar radio/mm observations \citep[e.g.,][]{2000ApJ...545.1058B,2013ApJ...777L..34S,2020ApJ...895...62S,2019MNRAS.484..648P}. This Sun-as-a-star approach has recently received increasing attention in the context of next-generation exoplanet observations for which it is crucial to separate observable exoplanet signatures from the host star’s “background radiation” 
\citep[see][and references therein]{2023RASTI...2..148R}.

\subsection{The solar-stellar connection}
\label{sec:stellar}
The Sun, which can be observed spatially resolved, serves as a fundamental reference case for studying the dynamic nature of stellar atmospheres and solar/stellar activity including the occurrence of flares \citep[e.g.,][see also Section~\ref{sec:flares}]{2015ApJ...814L..21D,2018A&A...612A..44S,2022ApJ...939...98O,2023AJ....166..173Z,2024ApJ...966...45M,2024A&A...682A..46P}. 
In addition to serendipitous detection as a by-product of other observations and surveys, dedicated stellar observations are essential  for understanding the structure and activity of stellar atmospheres in general. The sample by \citet{2021A&A...655A.113M} demonstrates the potential of stellar mm observations and, at the same time, illustrates the scarcity of suitable observations.  AtLAST could make substantial contributions in this regard, either through targeted observations or also through coincidental (or even archival) detections. 
In particular, observations of stellar flares in the (sub-)mm range are extremely rare \citep{MacGregor2018ApJ...855L...2M,MacGregor2020ApJ...891...80M}, and yet are crucial for unveiling the physics of these events including particle acceleration and plasma heating and for evaluating the impact of flares on the habitability of exoplanets \citep[see, e.g.,][and references therein]{2003ApJ...598L.121L,2010AsBio..10..751S,2022A&A...667A..15K,2023A&A...676A.139B}. 
At millimeter wavelengths, the flux density of the few detected stellar flares tends to be 10 - 1000 times higher than the quiescent flux. Assuming a typical quiet chromospheric temperature of $\sim10^5$\,K (at 100\,GHz) as seen in cool stars \citep{2021A&A...655A.113M}, the flux density during a flare can vary in the range $\sim$ 0.5 - 50\,mJy for a star at a distance of 10\,pc. 
Consequently, stellar flares can be detected even when the stellar disk is spatially unresolved. Observing a broad sample of stellar targets with varying flare intensity distributions therefore complements high-resolution solar flare observations. Translating solar observations into stellar signals, either through dedicated instruments like HARP-N \citep{2012SPIE.8446E..1VC,2015ApJ...814L..21D} or POET \citep{2023spfi.confE..12S}, or by processing, i.e. by employing Sun-as-a-star observational techniques, offers significant scientific potential. Such approaches contribute to the study of stellar atmospheres and can enhance our understanding of the physical mechanisms behind solar and stellar flares \citep[see, e.g., ][]{2022ApJ...939...98O,2024ApJ...966...45M,2024A&A...682A..46P}.

It is important to emphasise that the flux varies on timescales of a few seconds to minutes. 
Observing the temporal evolution of a flare, which is essential for exploring the physical mechanisms, thus requires time sequences with accordingly high temporal cadence. Such data sets are typically not generated from a serendipitous detection or as part of a survey but require a dedicated sit-and-stare observation of the same region in the sky. 
Should the required ultra-high temporal resolution not be achievable with full-disk scans of the Sun (see Section~\ref{sec:technical}), high-cadence flare observations of even a limited FOV could be used to transform the data into a Sun-as-a-star observation using a method like, e.g., the one by \citet{2023arXiv231106200P}.

AtLAST would be capable of such sit-and-stare observations, providing crucial time-resolved light curves for stellar flares. 
Given the high sensitivity of AtLAST, stellar flares in cool main sequence stars at distance of up to $\sim$50\,pc should be detectable with integration times of a few seconds, thus resolving the relevant timescales. In addition, the anticipated unique capability of observing at multiple frequencies simultaneously, which is typically not provided by other sub-mm instruments and surveys, especially when combined with polarisation capabilities, would enable AtLAST to utilise  essential diagnostics such as the variation of the mm spectral index, an indicator of stellar atmospheric activity \citep{2022A&A...664L...9M}, and strong constraints to the emission mechanism and estimates of the magnetic field strengths in the flaring active regions. 

The full potential of flare observations at millimeter wavelengths is only unlocked through coordinated, strictly co-simultaneous observations with other telescopes covering as large a wavelength range across the electromagnetic spectrum as possible. Such coordinated campaigns, which have become common in multi-messenger astronomy and are basically standard procedure for observations of the Sun, demand for PI-driven operations. 
The scarcity of such coordinated stellar observing campaigns is among the main reasons why there has been only a relatively small number of observations that are indicative of a stellar coronal mass ejection (CME) – an essential component of exo-space weather 
\citep{1990A&A...238..249H,1993A&A...274..245H,1994A&A...285..489G,1997A&A...321..803G,2004A&A...420.1079F,2016A&A...590A..11V,2019NatAs...3..742A,2021NatAs...5..697V,2022NatAs...6..241N}. 
The success of such campaigns would be greatly enhanced by AtLAST's large FOV and high angular resolution, thus being able to observe multiple stellar targets at the same time while separating the flare source(s) from other components such as close-by active companion stars or disks, making AtLAST ideal for time-resolved coordinated observations of stellar flares. 
See also the time-domain science cases for AtLAST  \citep{OREtimedomain}. 
  
\section{Observing the Sun with AtLAST} 
\label{sec:technical}

\subsection{Diagnostic capabilities and instrumental requirements}

There are two major constraints for solar observations at millimeter wavelengths: 
(i)~As the Sun is by far the brightest millimeter source in the sky, sensitivity is not an issue and very short integration times and thus even on-the-fly scanning is possible.
(ii)~
The Sun evolves on short time scales down to seconds and below depending on the scientific target, which prohibits long integration/scan times. Imaging of solar data thus needs to be done as snapshot imaging or over short time windows that should not exceed 1\,min for the expected resolution (on the Sun, temporal and spatial scales tend to correlate).
As will be detailed in this section, the combination of high angular resolution, high temporal cadence (or at least short integration time), wide frequency coverage, and large dynamic range (i.e., a high ratio between the maximum and minimum measurable brightness temperatures)  are crucial for meaningful solar observations with AtLAST \citep{Mroczkowski2024}.   In addition, full-Stokes polarimetry would greatly enhance AtLAST diagnostic potential for the Sun.

\subsubsection{High angular resolution and spatial mapping} 
\label{sec:spatialmap}

High spatial resolution is of utmost importance for solar observations as much of the relevant dynamics occurs on small spatial scales. Achieving sufficiently high resolution is challenging at millimeter wavelengths, in particular for single-dish telescopes such as AtLAST. While AtLAST's aperture sets a limit,  the achievable resolution is still in the relevant range and superior to any previous single-dish observation of the Sun. With an aperture of 50\,m, the achievable angular resolution would be 6.5'' at a wavelength of 1.3\,mm ($\sim$230\,GHz, ALMA Band~6) and down to just 1.5'' at 0.3\,mm ($\sim$950\,GHz, ALMA Band~10) (see Fig.~ \ref{fig:ifov}).

\paragraph{Fast on-the-fly scanning}~techniques (with a single beam) are used for creating maps of the whole solar disk or limited regions by, e.g, ALMA \citep{2017SoPh..292...88W}, and could be implemented for AtLAST, too \citep{kirkauneinprep}. 
The difference to mosaics is that the telescope is not consequently pointing at a set of locations on the Sun, possibly staying at each location for a few seconds as done for ALMA interferometric mosaics, but that telescope keeps observing while moving across the Sun to acquire a time stream that can then be converted into a map. 
This scanning motion could be accomplished by moving the primary or secondary reflector. 
Scanning the whole disk with an ALMA TP antenna in a double-circle pattern \citep{2015ASPC..499..347P} takes on the order of 10\,min. See Fig.~\ref{fig:almasdo} for examples. The roughly 4 times larger aperture of AtLAST compared to an ALMA 12-m antenna results in an accordingly much smaller ($\sim$6\,\%) beam area. 
Scanning the whole disk of the Sun would take considerably longer (on the order of ($(\frac{50\,\mathrm{m}}{12\,\mathrm{m}})^2 \approx 17$~times, i.e. $\sim$3~hours), which would be sufficient for synoptic studies with just one scan per day but prohibitive for science cases that rely on high temporal cadence. In the latter case, scans would then need to be limited to smaller region, although a compromise between size of the mapped region and overall cadence can be made. In either case, for a time series, the scanning could be continuous.

\paragraph{A multi-pixel detector}
would significantly enhance the diagnostic possibilities for solar observations with AtLAST. 
Even for a modest setup with a detector with 1000~pixels in a concentric setup with no overlap, spaced by one beam width, the resulting instantaneously covered region on the Sun (hereafter referred to as the instantaneous field of view, FOV) would cover a region with a radius of about 36~beams. While a single AtLAST beam would be a factor $50/12$ smaller than the beam of a 12\,m ALMA antenna, using 1000~pixels would effectively make the instantaneous FOV $36 \times \frac{12}{50} \approx 9$ times larger. 
As shown in Fig.~\ref{fig:ifov}, the diameter of the FOV would range from $\sim45$'' at 950\,GHz (ALMA Band~10) to $\sim450$'' at 100\,GHz (ALMA Band~3). 
Such a FOV would cover a complete sunspot even at the highest frequencies and whole Active Regions at the lower frequencies. The angular resolution of the final imaging products could be increased by means of a rapid small scan pattern (on scales comparable to the beam size and below) that compensates for the above-described pixel spacing.  
The diameter of the instantaneous FOV would be halved again for a denser pixel setup that allows for Nyquist sampling, which then would not require the mentioned scan pattern.   
Please see Sect.~\ref{sec:instrument} for a detailed discussion of the detector requirements for solar observations.

\paragraph{Multi-frequency synthesis} might offer means to further increase AtLAST angular resolution. However, as the formation height range of the mm continuum (see also Sect.~\ref{sec:coronalheating})  depends on the frequency, only a moderate frequency range should be considered. Depending on the receiver design,  multi-frequency synthesis could be an option that could be applied flexibly on AtLAST data sets.

\subsubsection{Frequency coverage and spectral setup}

An essential aspect to the continuum radiation is that the formation height range from where most of the emission emerges depends on the observing frequency or wavelength.  At the highest frequency accessible with ALMA, i.e. 950\,GHz (or a wavelength of 0.3\,mm), the continuum radiation stems from the upper photosphere and lower chromosphere, whereas the continuum radiation at the shortest frequency of 35\,GHz (equivalent to a wavelength just short of 1 cm) originates from the uppermost chromosphere with possible contributions from the transition region. Covering a large frequency range simultaneously or at least rapidly scanning through frequency has thus the potential to map a large height range in the solar atmosphere. In case of scanning through frequency, the scanning speed must be short compared to the dynamical timescales in the solar atmosphere, which are down to seconds and generally well below 1\,min (depending also on the achieved spatial resolution). 

Observations that cover extended regions on the Sun (or even the whole Sun) across a large frequency range would enable the reconstruction of the three-dimensional thermal structure of the solar chromosphere via tomographic and inversion techniques. Producing a time series at sufficiently high cadence would then enable to assess the temporal variation of the 3D structure --- a data product with the potential of truly ground-breaking scientific impact for a very large range of essential topics in solar physics. 

The observable frequency range should extend at least from 90\,GHz to 660\,GHz. 
Observing at higher frequencies of up to 1\,THz would provide higher  angular  resolution but adequate observing conditions may not occur every day, making these bands unreliable in the context of daily synoptic observations. 
Even for continuum observations, a high frequency resolution is desirable but already a frequency increment of $\sim 100$\,GHz would allow for excellent scientific results. 

\label{sec:spectrallines} 
While the aforementioned considerations apply to continuum observations, a large number of spectral channels with high spectral resolution and a flexible set-up of the receiver bands would offer additional possibilities. 
While it has yet to be seen if spectral lines, mostly hydrogen recombination lines and potentially CO, can be detected at millimeter wavelengths, they would have large diagnostic potential for assessing the thermal, kinetic and magnetic state of the chromospheric plasma. 
It is  difficult to answer right now which frequency setup would be necessary to exploit the diagnostic potential of spectral lines but, based on tentative simulations, hydrogen recombination lines might be spread out over one or a few GHz but would require nonetheless a spectral resolution better than 15.6\,MHz (as currently offered by ALMA) in order to exploit the full diagnostic potential (e.g., Doppler shifts for line-of-sight velocities).

\subsubsection{Sensitivity, integration time, dynamic range}

The Sun is by far the brightest millimeter source in the sky with typical brightness temperatures of several thousand Kelvin.  
Consequently, sensitivity is not a concern and the integration time can be very low, allowing for ultra-high cadence, which is of essence for a range of science cases. 
The integration time can be much below 1\,s depending on possible attenuation strategies (e.g., solar filter or detuning).  
Important for solar observations, however, is a large dynamic range.  
ALMA observations in Band~3 and 6 show a wide range of brightness temperatures from as low as 3000\,K to above 13\,000\,K \citep[see, e.g., the data sets in the Solar ALMA Science Archive,][]{2022A&A...659A..31H}. 
In general, the mean brightness temperature of the Sun depends on the frequency because the atmospheric height range of the mapped layer depends on the frequency and thus on the thermal stratification of the solar atmosphere. In principle, the highest ALMA receiver bands map the transition between the upper photosphere and lower chromosphere (i.e., the classical ``temperature minimum'' region) with comparatively low temperatures, whereas the lowest ALMA bands (i.e., at higher frequencies) map radiation emerging from the on average much hotter upper chromosphere close to the transition region with its steep jump in temperature.
See also \citet{2022ApJ...933..244H}  for a discussion of brightness temperatures obtained with ALMA in comparison to other (less temperature-sensitive) diagnostics and \citep{2019ApJ...877L..26L} for an observation of a particularly cool region. In addition to this already wide brightness temperature range, solar flares as a key science case for solar observations with AtLAST (Section~\ref{sec:flares}) define the upper end of extreme temperatures that occur on the Sun. 
Depending on the strength (i.e. overall energy release) of a flare, plasma temperatures exceed $10^6$\,K or even $10^7$\,K. 
A flare thus emits orders of magnitude more radiation than the average (quiescent) Sun as discussed above. It should be noted that the corresponding peak brightness temperature also depends on the observing frequency and the beam size, which is typically too large to resolve the source. From observations of a (very strong) X-class flare with the 1.5m-aperture Solar Submillimeter Telescope \citep[SST,][]{2008SPIE.7012E..0LK} 
at 212\,GHz an excess brightness temperature of \mbox{$1.75\times 10^6$\,K} for an assumed source size of 25'', which would remain unresolved \citep{2018SpWea..16.1261G}. In comparison, AtLAST would have an angular resolution of $\sim$7'' at 212\,GHz and would thus be able to resolve the source and thus brightness temperatures in excess of $10^6$\,K. At shorter wavelengths in the sub-mm range, we rely on simulations to estimate the expected brightness of flares (see Figure~\ref{fig:synthspectra}).

Addressing the full range of science cases would thus require to account for a brightness temperature range from typically 3000\,K up to 14\,000\,K for the non-flaring Sun, depending on frequency and resolution, but up to millions of K for flares.  Consequently, obtaining a large dynamic range for solar observations and at the same a high sensitivity for non-solar observations is a big challenge for designing of a receiver system that is intended to cover all science cases.

\subsubsection{Polarisation} 
\label{sec:polar}
Magnetic fields play an essential role in the Sun's structure, dynamics, and activity. Their impact is clearly visible in Active Regions most notably in the form of sunspots and as large-scale prominences but also the Sun's Quiet Regions harbour magnetic fields in various configurations and strengths. The latter becomes immediately obvious from high-resolution H$\alpha$ observations \citep[see, e.g.,][]{2008SoPh..251..533R,2023A&A...673A..11R}. 
High-resolution magnetograms obtained from the spectral lines in the photosphere (i.e., the Sun's surface layer) have become a staple in contemporary solar physics relevant to a large range of scientific topics but often also just for context. Reliably measuring the magnetic field in the solar chromosphere above, i.e. the layer that would be observed with AtLAST, is still a current technological challenge, again with much diagnostic potential  \citep[see, e.g., ][]{2020A&A...644A..43P,2021A&A...645A...1V,2022A&A...664A...8M}. 
The measured polarization state provides a measure for the longitudinal component of the magnetic field vector in the same layer as the continuum radiation. In the same way, a scan through wavelength (and thus through formation height) can be used to reconstruct the three-dimensional magnetic field structure in the solar chromosphere. Measuring the magnetic field in this layer is in itself a hot topic with potentially ground-breaking results. 

Early observations with ALMA at a frequency of 100\,GHz, which were obtained as science verification data in Band~3 (ADS/JAO.ALMA\#2011.0.00011.E), indicate that the circular polarization degree is at least 1\% . This value is considered a minimum, as the observations were carried out with the 12m-Array only \citep{2024arXiv240106343S}. 
First regular ALMA observations of the Sun with full polarisation in Band~3 were carried out in 2024 (Cycle 10). 
Important lessons are expected from these observations, which  would have implications for full polarisation observations with AtLAST. 
In general, circular polarization levels of up to 5\% are expected for the Sun at mm wavelengths. 

\subsection{Observing modes} 

In order to address the aforementioned key science cases (see Sect.~\ref{sec:keyscience}), it would be best to combine 1)~long-term synoptic observations with a very low daily load on observing time with 2)~campaign-based observations. 
The synoptic observations could consist of one daily full-disk map or preferably a 
10-30\,min long time sequence of full-disk maps across a wide predefined frequency range. The resulting long-term data set would be of essential value for studying the solar cycle and all related variations imprinted in the Sun's chromosphere with fundamental implications for stellar activity cycles and their impact on the habitability of exoplanets \citep{2023BAAS...55c.428W}. The campaign-based observations would be more adjusted towards the requirements of individual science cases. The scheduling for observations that require the existence of an Active Region could be planned with just a few days ahead as the co-ordination with other space-borne and ground-based observatories would boost the scientific impact of such AtLAST data. Alternatively, a ToO Target-of-Opportunity (ToO) mode (also possibly confined to a campaign) would ensure observations of flares whenever a suitable Active Region appears.

\subsubsection{Solar observing mode 1 -- Synoptic full-disk scans} 
\label{sec:synoptic}

A field-of-view (FOV) that would cover the whole Sun plus a margin around, which is important to include the Sun's outer atmosphere, has a diameter of $\sim 2000$''. Scanning such a region with the comparatively small beam size of AtLAST would require long scan times (see Sect. ~\ref{sec:spatialmap}). 
Therefore we expect that multi-pixel detectors will be needed for solar disk mapping as well as other non-solar large FOV studies. The corresponding time needed for scanning the whole FOV would be reduced basically by the number of simultaneously used beams (pixels) and thus the resulting size of the instantaneously covered FOV (see also Sect.~\ref{sec:spatialmap}). 
The required scan time for the whole solar disk would scale accordingly with respect to the single-dish Total Power (TP) scans with ALMA antennas, which take about $\sim10$\,min. Consequently, a full-disk scan at the lower frequencies  could potentially be completed in a few minutes or less, aided by the anticipated high slew speed of 3\,deg/s. 
A smaller number of pixels, e.g. $\sim 250$~pixels, would directly affect the covered FOV, resulting in correspondingly longer scan times.  
Please note that very short integration times are sufficient as the Sun is the brightest mm source on the sky, and correspondingly mapping speed is not at all dependent on sensitivity requirements, rather it depends solely on the receiver FOV and antenna slew rates.  
See Sect.~\ref{sec:instrument} for more considerations regarding multi-pixel detectors. 

The short times for completing a scan of the whole solar disk would allow for scheduling a daily solar map, which would have enormous potential for studying the long-term evolution and the solar activity cycle (see Sect.~\ref{sec:solcycle}).

\begin{figure*}[!t]
\vspace{-3mm}
\centering
\includegraphics[width=16cm]{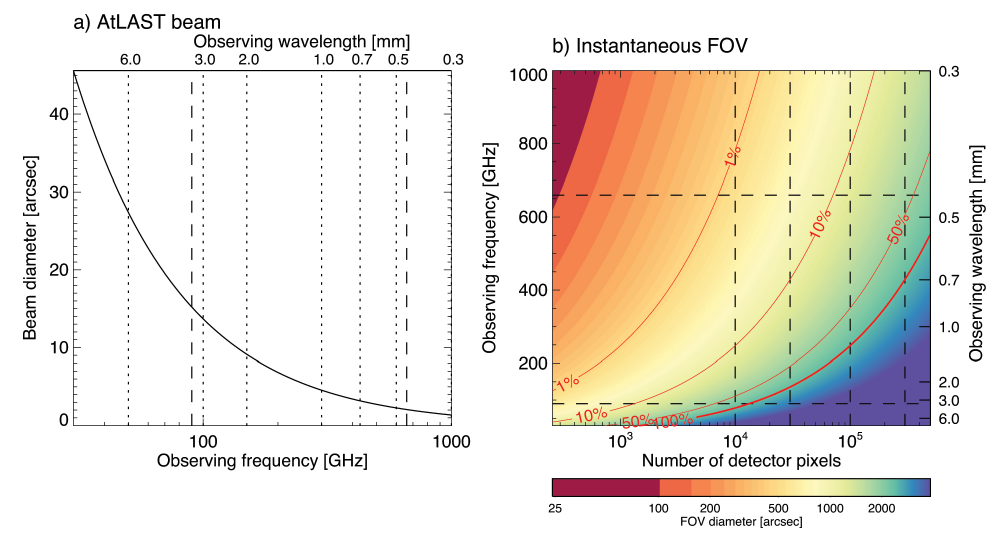}
\vspace{-2mm}
 \caption{\textit{Left:} AtLAST beam diameter as a function of observing frequency. The vertical dashed lines mark frequencies of 90\,GHz and 660\,GHz, which we here define as a minimum requirement for the frequency range. 
\textit{Right:} The resulting diameter of the instantaneous FOV as a function of number of detector pixels and observing frequency (see color legend). The resulting ratio of the FOV area and the whole disk of the Sun is overlaid as red contours. The dashed lines mark selected frequencies and detector numbers for reference. 
The pixels are assumed to have the size of the corresponding beam and are packed into a circle with no overlap. See Sect.~\ref{sec:instrument} for details.
} 
\label{fig:ifov}
\end{figure*}

\subsubsection{Solar observing mode 2 -- High-resolution time-dependent observations}
A selected target, e.g. a sunspot, would be followed in time for one or several hours (compensating for motion on the sky plus solar rotation) and observed with preferably high cadence ($\sim$1\,s). For comparison, ALMA currently observes the Sun at 1\,s cadence, demonstrating the scientific potential of such a high cadence, e.g. for the study of waves \citep[e.g., ][]{2020A&A...634A..86P,2021RSPTA.37900185E,2021RSPTA.37900184G,2021RSPTA.37900174J}. 
This mode, which is equivalent to observations at visible wavelengths in a limited field of view, is suitable for a large range of science cases that require a combination of high spatial and temporal resolution. 

\subsubsection{Solar observing mode 3 -- Regional scan sequences} 
As a compromise between full-disk scans (mode~1) and single-target tracking (mode~2), AtLAST could perform scans of extended regions (or alternatively mosaics, see Section~\ref{sec:spatialmap}). 
The size of the covered region on the Sun would depend on the instantaneous field of view of the multi-pixel detector (see Section~\ref{sec:instrument}) and the maximum acceptable temporal cadence. Such scans could be repeated continuously for some time, e.g., one hour, providing a good temporal cadence of a large field of view. This mode could be useful for observations of large-scale features such as filaments or entire Active Regions. Such observations would facilitate the detection of flares and filament eruptions and, at the same time, capture their evolution at adequate temporal resolution.

\subsection{Instrumentation} 
\label{sec:instrument}

The detectors described in this section assume technological progress on a time scale of about 10~years. The specifications are provided by the AtLAST instrument working group\footnote{\url{https://www.atlast.uio.no/memo-series/memo-public/instrumentationwgmemo4_29feb2024.pdf}}. 
Please note the proposed instruments would have different operation modes and would also be suited for slightly different science cases as the different setups have to make a compromise between instantaneously covered FOV, spectral resolution, frequency range, and possibly temporal cadence due to costs and technological limitations. 
For instance, an instrument with a too small FOV due to a limited number of pixels would make it more challenging to capture flares, whereas limiting to one frequency band (with spectral channels or not) would only permit observations of the mm continuum originating from a limited height range in the solar atmosphere. In contrast, a multi-chroic setup could facilitate the reconstruction of the atmospheric 3D structure but at the expense of spectral resolution. Consequently, prioritising one of these properties will come at the cost of the others, warranting a detailed study to determine the best instrumental setup in view of the science cases described in Section~\ref{sec:keyscience}. 
In Figure~\ref{fig:ifov}, the instantaneously covered FOV of multi-pixel detectors with $n$~pixels is illustrated. The pixels are assumed to have the same diameter as the beam diameter $d_\mathrm{beam}$ at a given observing frequency and are packed with no overlap.  
Treated as a circle packing problem, the resulting diameter of the FOV is approximated as 
\begin{equation} 
d_\mathrm{fov} \approx  \sqrt{\frac{2\ \pi}{3 \sqrt{3}}} \ \  d_\mathrm{beam}  \sqrt{n} \ . 
\end{equation}

\subsubsection{Multi-chroic continuum camera}
\label{sec:multichroic}
A first-generation multi-chroic camera for continuum observations could have 30\,000~pixels covering 3 frequencies simultaneously while a second-generation camera could be extended to possibly 300\,000~pixels covering 6 frequencies simultaneously. 
The frequencies would be preferably spread equally over the range from $\sim100$\,GHz to 700\,GHz but adjusted according to the atmospheric transmission at the AtLAST site\footnote{\url{https://almascience.eso.org/about-alma/alma-site}}, e.g., 100, 220, 340, 460, 615, and 700\,GHz. 
Full polarisation capability would substantially enhance the diagnostic possibilities (see Sect.~\ref{sec:polar}). 

The details of the detector setup have to be investigated in detail. That includes the spacing between pixels, the total number of pixels, and the resulting instantaneously covered FOV. 
A fast small circular scanning pattern could be used to fill intermediate pointings.   Combining data from circular scans (if possible on a second time scale or at least <10\,s) would then allow to produce data with higher angular resolution. 
Potential scanning strategies are currently being investigated  \citep{kirkauneinprep}. 

\subsubsection{Multi-pixel heterodyne receiver system}
\label{sec:heterodyne}

A full heterodyne system with 64~spectral pixels (spaxels) and 16\,GHz bandwidth seems technologically feasible, whereas, on a 10-years perspective, systems with at least 256~spectral pixels (spaxels) and 30\,GHz bandwidth might become possible. 
Such a detector would allow for exploiting the spectral domain but only with a much smaller number of detector elements as compared to multi-chroic continuum cameras due to the higher cost per pixel (see Sect.~\ref{sec:multichroic}), strongly limiting the instantaneously covered region on the Sun. In combination with spatial scanning strategies, multi-pixel heterodyne receiver system would require compromises between covered field of view and overall temporal cadence, which would need to be adjusted to the needs of individual science cases. These limitations would only make sense if spectral lines are discovered on the Sun which would justify these limitations for accessing the diagnostic potential of spectral lines (see Sect.~\ref{sec:spectrallines}).

\subsubsection{Integrated Field Unit}
\label{sec:ifu}

While multi-chroic continuum cameras (see Sect.~\ref{sec:multichroic}) and heterodyne receiver systems (see Sect.~\ref{sec:heterodyne}) are the extreme cases considered here in terms of field of view versus spectral resolution, Integrated Field Unit could provide a compromise. First-generation units are expected to have a spectral resolving power of $R = \lambda/d\lambda = 500$ with 10\,000 detectors (number of spaxels $\times$  number of channels) and a frequency range of 70-690\,GHz, second generation units could reach $R=2000$ with 50\,000 detectors covering an even wider frequency range. 
While a resolving power of $R = 2000$ would open diagnostic possibilities for spectral line observations in a narrow frequency range, it would only make sense if spectral lines at mm wavelengths were observable in the Sun.  
Integrated Field Units would be less efficient than multi-chroic continuum cameras for observing a large frequency range and thus a large height range in the solar atmosphere due to the still limited number of spaxels. With even only 10~frequencies, only 5000~spatial pixels would be available, which is considerably lower than what is expected for multi-chroic cameras. As for multi-pixel heterodyne receiver systems, IFUs would likely only make sense if spectral lines can be observationally exploited. In any case, a realistic evaluation of the scientific potential of IFUs will require a dedicated study.

\subsection{Synergies for development and co-observing}
\label{sec:synergy}

Given the complex and dynamic nature of the solar atmosphere, coordinated multi-wavelength multi-instrument observing campaigns are the default modus operandi. This typically involves ground-based and space-borne observatories, which provide complementary data exploiting different continua and spectral lines across the whole spectrum, thus providing large data sets that probe the (ideally complete) thermodynamic and magnetic state of the solar plasma across preferably all atmospheric layers. While such coordinated campaigns are subject to the different time zones of participating instruments, there are many successful examples including ALMA observations of the Sun, which are typically complemented with space-borne observations with the Solar Dynamics Observatory (SDO) and the Interface Region Imaging Spectrograph (IRIS). See the Solar ALMA Science Archive \citep[SALSA,][]{2022A&A...659A..31H} for examples. 

Solar observations with AtLAST will be no different. The idea behind the suggested synoptic full-disk observations (see Sect.~\ref{sec:synoptic}) is to provide a reference for complementary observations and to become a cornerstone of future solar observations very much like SDO today. 

The development of AtLAST can be fundamental to cover key ranges of the sub-mm spectrum for joint flare observations, along with instruments already in operation, such as the 30~THz cameras at Mackenzie University in São Paulo and at CASLEO, Argentina \citep{BR30T2015SoPh..290.2373K}, the new High Altitude THz Solar photometer at 15~THz \citep[HATS, ][]{HATS2020SoPh..295...56G} and future telescopes, such as the Solar Submillimeter Telescope Next Generation \citep[SSTng, ][]{SSTng10265358}.


\section{Summary of telescope requirements} 

AtLAST \citep{Mroczkowski2024} observations of the Sun would produce a large range of valuable contributions to science cases that are difficult if not impossible to address with other existing or planned observatories. 
The need of high temporal, spatial, and to some extent spectral resolution makes adequate solar observations challenging. 
At the same time, resolution in one dimension can be sacrificed to boost the resolution in another dimension, e.g., lowering the temporal cadence for increased spatial resolution. 
Consequently, the best approach combines a multi-pixel detector with fast on-the-fly scanning. Appropriate imaging strategies could be developed during the commissioning phase and  through simulations at an earlier stage. 
Additionally, in case that spectral lines at mm wavelengths are discovered in the Sun, heterodyne receiver systems of IFUs might have important scientific applications but require dedicated studies to realistically assess the feasibility and applicability given the limited possible detector sizes. 
Since the Sun is a very bright mm source, it could even be possible to split the optical path and feed multiple instruments that are optimised for different, complementary  diagnostic purposes.   
Also, it has yet to be seen which solar science cases  require instruments specifically designed for solar observing and which cases could be addressed with general instruments.  

To fully unlock AtLAST's potential in this respect, the following instrumental properties are essential: 

\begin{itemize} 
\setlength{\itemsep}{0mm}
\item \textbf{Sufficient instantaneous field-of-view:} A multi-pixel detector is a fundamental requirement as the instantaneously covered region on the sky with a single beam would be insufficient for most solar science cases. 
\item \textbf{Wide frequency coverage:} 
A wide (quasi-)simultaneous frequency coverage beyond the current setup of ALMA's receiver bands, possibly even covering the whole frequency range from (at least) 90\,GHz to 660\,GHz, has large scientific potential as it facilitates simultaneous mapping of an extended height range in the solar atmosphere. A multi-chroic camera with a large number of pixels seems to be a promising choice for continuum observations. 
\item \textbf{Full polarisation capabilities} are desirable in order to provide information on chromospheric magnetic fields. 
\item \textbf{High temporal cadence:} 
While integration times are very short for a bright object like the Sun, spatial (or spectral) scanning in order to cover an adequate spatial (or spectral) region is constrained by the short dynamic timescales. Preferably, the temporal cadence should be less than 1\,min but the shorter the better. High detector readout rates are required, especially for fast scanning. 
\item \textbf{Adequate brightness temperature range and accuracy:}
As the Sun is the brightest mm source in the sky and can vary in brightness significantly, in particular during flares, a solar filter with well-defined characteristics and large dynamic range for the detector are needed. 
A brightness temperature accuracy of 100\,K or better is important for solar science cases. Consequently, the need for detector cooling will depend on the achievable signal-to-noise ratio.   
\item \textbf{Spectral resolution:} 
Spectral lines -- if successfully detected in the Sun -- would enormously increase the diagnostic capabilities. Designing a suitable instrument, possibly based on Integrated Field Units, would require a dedicated study. 
\end{itemize}

\section*{Grant information}
This project has received funding from the European Union’s Horizon 2020 research and innovation programme under grant agreement No.~951815 (AtLAST). 
In addition, we acknowledge support by the Research Council of Norway through the EMISSA project (project number 286853) and the Centres of Excellence scheme, project number 262622 (``Rosseland Centre for Solar Physics''). 
GF acknowledges support from NSF grants AGS-2121632 and AST-2206424 and NASA grant 80NSSC23K0090.
FM acknowledges financial support from grants \#2022/12024-0 and \#2013/10559-5, São Paulo Research Foundation (FAPESP).
L.D.M. acknowledges support by the French government, through the UCA\textsuperscript{J.E.D.I.} Investments in the Future project managed by the National Research Agency (ANR) with the reference number ANR-15-IDEX-01.
M.L. acknowledges support from the European Union’s Horizon Europe research and innovation programme under the Marie Sk\l odowska-Curie grant agreement No 101107795.
PJAS acknowledges support from Conselho Nacional de Desenvolvimento Científico e Tecnológico (CNPq) (contract 305808/2022-2), Fundo Mackenzie de Pesquisa e Inovação (MackPesquisa) project 231017 and Fundação de Amparo à Pesquisa do Estado de São Paulo (FAPESP) contract 2022/15700-7.

\section*{Software}
The calculations used to derive
 integration times for this paper were done using the AtLAST sensitivity calculator, a deliverable of Horizon 2020 research project `Towards AtLAST', and available from \href{https://github.com/ukatc/AtLAST_sensitivity_calculator}{this link}.

\section*{Acknowledgements}
This article, which was composed by the Working Group for Solar Science with AtLAST (SoSA) as part of the AtLAST design study, makes use of the following ALMA data: 
\linebreak
ADS/JAO.ALMA\#2011.00020.SV, 
\linebreak
ADS/JAO.ALMA\#2017.1.00009.S. 
\linebreak
ALMA is a partnership of ESO (representing its member states), NSF (USA) and NINS (Japan), together with NRC (Canada) and NSC and ASIAA (Taiwan), and KAS (Republic of Korea), in cooperation with the Republic of Chile. The Joint ALMA Observatory is operated by ESO, AUI/NRAO and NAOJ. 
\bibliographystyle{apj_mod}

\end{document}